\documentclass[conference]{IEEEtran}
\IEEEoverridecommandlockouts
\usepackage{cite}
\usepackage{amsmath,amssymb,amsfonts}
\usepackage{algorithmic}
\usepackage{graphicx}
\usepackage{textcomp}
\usepackage{xcolor}
\usepackage{hyperref}
\usepackage{longtable}
\usepackage{tabularx}
\usepackage{tcolorbox}
\usepackage{array}
\usepackage{booktabs}
\usepackage{xurl}
\usepackage{tablefootnote} 
\tcbuselibrary{skins}

\newtcolorbox{userbox}{colback=blue!5,colframe=blue!40!black,title=User}
\newtcolorbox{assistantbox}{colback=gray!10,colframe=black,title=Assistant}

\def\BibTeX{{\rm B\kern-.05em{\sc i\kern-.025em b}\kern-.08em
    T\kern-.1667em\lower.7ex\hbox{E}\kern-.125emX}}
\begin{document}

\title{HASHIRU: Hierarchical Agent System for Hybrid Intelligent Resource Utilization}

\author{
  \IEEEauthorblockN{Kunal Pai\IEEEauthorrefmark{1}}
  \IEEEauthorblockA{\textit{UC Davis} \\
    kunpai@ucdavis.edu}
  \and
  \IEEEauthorblockN{Parth Shah\IEEEauthorrefmark{1}}
  \IEEEauthorblockA{\textit{Independent Researcher} \\
    helloparthshah@gmail.com}
  \and
  \IEEEauthorblockN{Harshil Patel}
  \IEEEauthorblockA{\textit{UC Davis} \\
    hpppatel@ucdavis.edu}
  \thanks{\IEEEauthorrefmark{1}These authors contributed equally to this work.}
}
\maketitle
\begin{abstract}
Rapid Large Language Model (LLM) advancements are fueling autonomous Multi-Agent System (MAS) development. However, current frameworks often lack flexibility, resource awareness, model diversity, and autonomous tool creation. This paper introduces HASHIRU (Hierarchical Agent System for Hybrid Intelligent Resource Utilization), a novel MAS framework enhancing flexibility, resource efficiency, and adaptability. HASHIRU features a ``CEO'' agent dynamically managing specialized ``employee'' agents, instantiated based on task needs and resource constraints (cost, memory). Its hybrid intelligence prioritizes smaller, local LLMs (often via Ollama) while flexibly using external APIs and larger models when necessary. An economic model with hiring/firing costs promotes team stability and efficient resource allocation. The system also includes autonomous API tool creation and a memory function. Evaluations on tasks like academic paper review (58\% success), safety assessments (100\% on a JailbreakBench subset), and complex reasoning (outperforming Gemini 2.0 Flash on GSM8K: 96\% vs. 61\%; JEEBench: 80\% vs. 68.3\%; SVAMP: 92\% vs. 84\%) demonstrate HASHIRU's capabilities. Case studies illustrate its self-improvement via autonomous cost model generation, tool integration, and budget management. HASHIRU offers a promising approach for more robust, efficient, and adaptable MAS through dynamic hierarchical control, resource-aware hybrid intelligence, and autonomous functional extension.
Source code and benchmarks are available at \href{https://github.com/HASHIRU-AI/HASHIRU}{HASHIRU} and \href{https://github.com/HASHIRU-AI/HASHIRUBench}{HASHIRUBench}, and a \href{https://hashiruagentx-hashiruai.hf.space}{live demo} is available upon request.
\end{abstract}

\section{Introduction}\label{sec:introduction}

Rapid Large Language Model (LLM) advancements are reshaping AI, enabling complex language understanding, generation, reasoning, and planning \cite{brown2020language, devlin2019bert, raffel2020exploring}. This progress fuels the development of autonomous Multi-Agent Systems (MAS) where collaborative teams tackle problems beyond individual agent capabilities \cite{dorri2018multi, wooldridge2009introduction}. Collaborative MAS show potential in scientific discovery \cite{boiko2023emergent}, software engineering \cite{qian2023communicative}, data analysis, and decision-making \cite{wang2023decision}. The increasing complexity of tasks, evidenced by benchmarks requiring advanced reasoning (e.g., GSM8K \cite{cobbe2021gsm8k}, SVAMP \cite{patel2021svamp}), coding \cite{chen2021codex, pai2024codocbench}, and graduate-level knowledge \cite{phan2025humanitysexam}, necessitates agentic systems that effectively coordinate diverse cognitive resources \cite{wen2024benchmarkingcomplexinstructionfollowingmultiple}.

Despite this potential, contemporary agentic frameworks exhibit limitations: \textbf{rigidity} due to predefined roles hindering adaptation \cite{zhang2023building}; \textbf{resource obliviousness}, lacking mechanisms to optimize computational resources (API costs, memory, CPU), leading to inefficiency, especially with costly proprietary LLMs \cite{park2023generative}; \textbf{model homogeneity}, defaulting to a single powerful LLM and missing efficiency gains from diverse, smaller, or local models \cite{zhou2023agents}; and limited autonomous \textbf{tool creation and integration}, restricting dynamic self-improvement \cite{wang2023voyager, yao2022react, parisi2022talm}.

To address these challenges, we introduce \textbf{HASHIRU (Hierarchical Agent System for Hybrid Intelligent Resource Utilization)}, a novel MAS framework enhancing flexibility, resource efficiency, and adaptability. HASHIRU uses a hierarchical structure with a ``CEO'' agent dynamically managing specialized ``employee'' agents, instantiated on demand. Its \textbf{hybrid intelligence} strategically prioritizes smaller, local LLMs (e.g., 3B--7B, often via Ollama \cite{ollama}) for cost-effectiveness, flexibly integrating external APIs and larger models when justified by task complexity and resource availability under CEO management.

The primary contributions are:
\begin{enumerate}
    \item A novel MAS architecture with \textbf{hierarchical control} and \textbf{dynamic, resource-aware agent lifecycle management} (hiring/firing) governed by budget constraints (cost, memory) and an economic model discouraging excessive churn.
    \item A \textbf{hybrid intelligence model} prioritizing cost-effective, local LLMs while adaptively incorporating external APIs and larger models, optimizing the efficiency-capability trade-off.
    \item Integrated \textbf{autonomous API tool creation} for dynamic functional extension.
    \item An \textbf{economic model} (hiring/invocation fees) promoting efficient resource allocation and usage based on task needs and system constraints.
\end{enumerate}

This paper details HASHIRU's design. Section \ref{sec:background} discusses related work. Section \ref{sec:architecture} elaborates on the architecture. Section \ref{sec:casestudies} presents case studies demonstrating self-improvement capabilities. Section \ref{sec:experiments} describes the experimental setup and evaluation metrics. Section \ref{sec:results} reports results, and Section \ref{sec:limitations_future_work} concludes with limitations and future work.

\section{Background and Related Work} \label{sec:background}

Intelligent agent concepts have evolved from early symbolic AI \cite{russell2010artificial, shoham1994agent} to LLM-dominated frameworks leveraging models for reasoning, planning, and interaction \cite{wang2023survey, xi2023rise}. HASHIRU builds on this, addressing current limitations.

\textbf{Agent Architectures:}
MAS architectures vary, including flat, federated, and hierarchical \cite{dorri2018multi, horling2004survey}. Hierarchical models offer clear control and task decomposition but risk bottlenecks and rigidity \cite{gaston2005agenta,gaston2005agentb}. HASHIRU uses a \textbf{CEO-Employee hierarchy} for centralized coordination but distinguishes itself through \textbf{dynamic team composition}. Unlike systems with static hierarchies or predefined roles (e.g., CrewAI \cite{crewai}, ChatDev \cite{qian2023communicative}), HASHIRU's CEO dynamically manages the employee pool based on runtime needs and resource constraints.

\textbf{Dynamic Agent Lifecycle Management:}
Dynamic MAS composition is crucial for complex environments \cite{valckenaers2005trends}. Agent creation/deletion triggers often relate to task structure or environmental changes. HASHIRU introduces a specific mechanism where the CEO makes \textbf{hiring and firing decisions} based on a cost-benefit analysis considering agent performance, operational costs (API fees, inferred compute), memory footprint (tracked explicitly as a percentage of available resources), and concurrency limits. HASHIRU also incorporates an \textbf{economic model} with explicit ``starting bonus'' (hiring) and ``invocation'' (usage) costs. This economic friction aims to prevent excessive initialization or usage for marginal gains and promote team stability, a nuance often missing in simpler dynamic strategies.

\textbf{Resource Management and Agent Economies:}
Resource awareness is critical for scalable MAS. Economic research explores mechanisms like market-based auctions or contract nets for allocation \cite{clearwater1996market}. HASHIRU implements a more \textbf{centralized, budget-constrained resource management model}. The CEO operates within defined limits for financial cost and memory usage (as a percentage of total allocated). This direct management, particularly focusing on memory percentage, suggests practicality for deployment on local or edge devices with finite resources, contrasting with cloud systems assuming elastic resources \cite{park2023generative}. Frameworks like AutoGen \cite{wu2023autogen} and LangGraph \cite{langgraph} typically rely on implicit cost tracking without explicit multi-dimensional budgeting and control.

\textbf{Hybrid Intelligence and Heterogeneous Models:}
Leveraging diverse LLMs with varying capabilities, costs, latencies, and self-evolution is an emerging trend \cite{zhou2023agents,liang2024self}. Techniques like model routing select optimal models for sub-tasks. HASHIRU embraces \textbf{model heterogeneity} with a strategic focus: \textbf{prioritizing smaller (3B--7B), locally-run models via Ollama integration} \cite{ollama}. This emphasizes cost-efficiency, low latency, and potential privacy over systems defaulting to large proprietary cloud APIs (e.g., GPT-4 \cite{openai2023gpt4}, Claude 3 \cite{anthropic2024claude}). While integrating external APIs (potentially larger models), HASHIRU's default stance represents a distinct capability vs. efficiency balance.

\textbf{Tool Use and Autonomous Tool Creation:}
Tool use (APIs, functions) is fundamental for modern agents \cite{yao2022react, openai_func_calling}. Most systems use predefined tools. HASHIRU advances this with \textbf{integrated, autonomous API tool creation}. When needed functionality is missing, the CEO can commission the generation (potentially via a specialized agent) and deployment of a new API tool within the HASHIRU ecosystem. This self-extension capability differentiates HASHIRU from systems limited to static toolsets, moving towards greater autonomy and adaptability \cite{wang2023voyager, park2023generative}.

In summary, HASHIRU integrates hierarchical control, dynamic MAS, resource management, and tool use. Its novelty lies in the synergistic combination of: (1) dynamic, resource-aware hierarchical management with (2) an economic model for stability, (3) a local-first hybrid intelligence strategy, and (4) integrated autonomous tool creation. This targets key limitations in current systems regarding efficiency, adaptability, cost, and autonomy.

\section{HASHIRU System Architecture}
\label{sec:architecture}

HASHIRU's architecture addresses rigidity, resource obliviousness, and limited adaptability through a hierarchical, dynamically managed MAS optimized for hybrid resource utilization.

\subsection{Overview}
\begin{figure}[ht]
    \centering
    \includegraphics[width=0.45\textwidth]{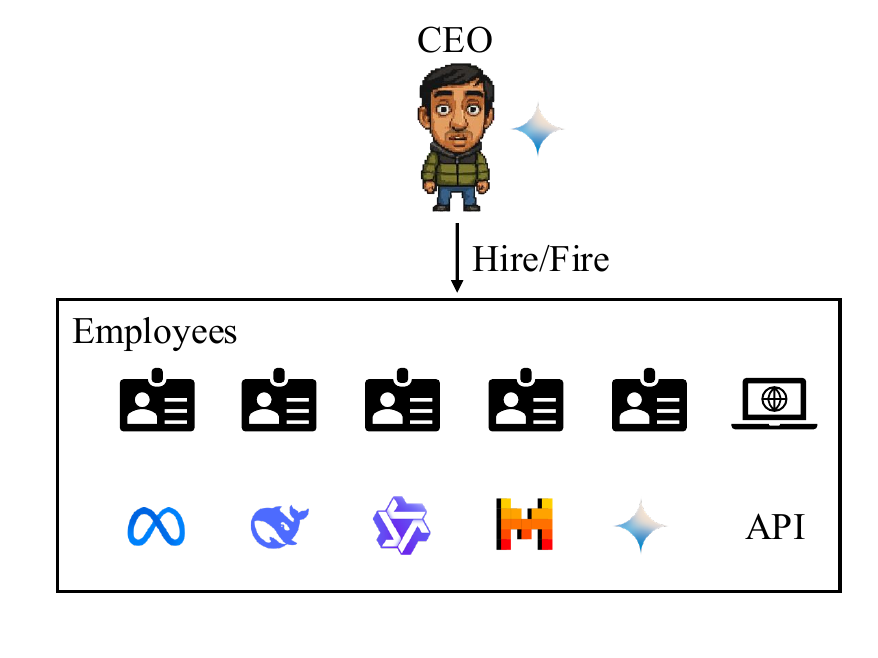}
    \caption{High-level architecture of the HASHIRU system, illustrating the CEO-Employee hierarchy.}
    \label{fig:arch}
\end{figure}

HASHIRU operates with a central ``CEO'' agent coordinating specialized ``Employees''. Key tenets:
\begin{itemize}
    \item \textbf{Dynamic Hierarchical Coordination:} CEO manages strategy, task allocation, and dynamic team composition.
    \item \textbf{Dynamic Lifecycle Management:} Employees are hired/fired based on runtime needs and resource constraints, governed by an economic model.
    \item \textbf{Hybrid Intelligence:} Strategic preference for LLMs within a predefined budget, while accessing external APIs/models.
    \item \textbf{Explicit Resource Management:} Continuous monitoring and control of costs against budgets.
    \item \textbf{Adaptive Tooling:} Using predefined tools alongside autonomous creation of new API tools.
\end{itemize}
Figure \ref{fig:arch} illustrates the structure.

\subsection{Hierarchical Structure: CEO and Employee Agents}
The system uses a two-tiered hierarchy:

\begin{itemize}
    \item \textbf{CEO Agent:} Singleton, central coordinator and entry point. Responsibilities:
        \begin{itemize}
            \item Interpreting user query/task.
            \item Decomposing main task into sub-tasks.
            \item Identifying required capabilities.
            \item Managing Employee pool (Section \ref{subsec:dynamic_mgmt}).
            \item Assigning sub-tasks to active Employees.
            \item Monitoring Employee progress/performance.
            \item Synthesizing Employee results into final output.
            \item Managing overall resource budget (Section \ref{subsec:resource_mgmt}).
            \item Initiating new tool creation (Section \ref{subsec:tooling}).
        \end{itemize}
        We use Gemini 2.0 Flash~\cite{gemini20flash} as the CEO agent. To further enhance its planning and reasoning abilities, its system prompt is designed to evoke inherent chain-of-thought processes~\cite{wei2022chain} when tackling complex user queries and managing sub-tasks. This complements its strong baseline reasoning capabilities, tool usage support, and cost efficiency, making it a practical and capable choice for our deployment.
    \item \textbf{Employee Agents:} Specialized agents instantiated by the CEO for specific sub-tasks. Each typically wraps an LLM (local via Ollama \cite{ollama} or external API) or provides tool access. Characteristics:
        \begin{itemize}
            \item Specialization: Capabilities tailored to task types (code, data analysis, info retrieval).
            \item Dynamic Existence: Created/destroyed by CEO based on need/performance.
            \item Task Execution: Receive task, execute, return result.
            \item Resource Consumption: Associated costs (API, hardware utilization) tracked by system.
        \end{itemize}
        Specialized employee agents are constructed using base models such as Mistral~7B~\cite{jiang2023mistral}, Llama~3~\cite{llama3herd}, Gemini~1.5~\cite{gemini1.5_report}, Qwen2.5~\cite{qwen2.5_report}, and DeepSeek-R1~\cite{deepseekr1_report}, with the CEO agent configuring them via tailored system prompts that it generates based on the task requirements.
        The models will be run locally using Ollama~\cite{ollama}, and via API calls to external models such as Gemini 2.5 Flash~\cite{gemini25flash}, Qwen QwQ~\cite{QwenQwQ32B2025}, Llama 4~\cite{Llama4Herd2025}, Mistral Saba~\cite{MistralSaba2025}, Hermes3~\cite{teknium2024hermes} and other models hosted on Hugging Face~\cite{huggingface2025}, Groq~\cite{groq2025}, Lambda.ai~\cite{lambda2025}, and other platforms.
\end{itemize}
This hierarchy facilitates task decomposition and result aggregation; the dynamic pool provides flexibility.

\subsection{Dynamic Agent Lifecycle Management}
\label{subsec:dynamic_mgmt}
A core innovation is the CEO's dynamic management (hiring/firing) of Employee agents. Driven by cost-benefit analysis, this optimizes task performance within resource constraints.

When a sub-task needs unavailable or inefficiently provided capabilities, the CEO may hire a new agent. Conversely, if an agent underperforms, is idle, costly, or resource limits are neared, the CEO may fire it. Decision factors:
\begin{itemize}
    \item \textbf{Task Requirements:} Needed capabilities for pending sub-tasks.
    \item \textbf{Agent Performance:} Historical success, output quality, efficiency.
    \item \textbf{Operational Costs:} API, estimated compute, or other costs.
\end{itemize}

HASHIRU includes an \textbf{economic model}:
\begin{itemize}
    \item \textbf{Hiring Cost (``Starting Bonus''):} A one-time cost incurred upon instantiation of local models, representing setup overhead. This cost can be quantitatively scaled based on the resource profile of the model (e.g., higher for models requiring more VRAM or complex setup).
    \item \textbf{Invocation Cost (``Salary''):} A recurring cost applied each time a local model is used, reflecting the operational load (e.g., inferred compute, system resource engagement). This abstracts the cost of utilizing local resources for a given task.
    \item \textbf{Expense Cost:} A recurring cost for external API calls (e.g., OpenAI, Anthropic), typically calculated based on token usage as per the API provider's documented pricing.
\end{itemize}
These transaction costs discourage excessive churn, promoting stability. The CEO evaluates if replacing an agent benefits outweigh hiring/firing costs plus operational differences. This combats rigidity and allows adaptation while managing budgets and preventing wasteful turnover.

\subsection{Hybrid Intelligence and Model Management}
HASHIRU is designed for \textbf{hybrid intelligence}, leveraging diverse cognitive resources. It strategically prioritizes smaller (3B--7B), cost-effective local LLMs via Ollama \cite{ollama}. This enhances efficiency, reduces external API reliance, and potentially improves privacy/latency.

The system also integrates:
\begin{itemize}
    \item \textbf{External LLM APIs:} Access to powerful LLMs (Gemini 2.5 Flash~\cite{gemini25flash}, etc.) when necessary, subject to cost-benefit.
    \item \textbf{External Tool APIs:} Third-party software/data source integration.
    \item \textbf{Self-Created APIs:} Tools generated by HASHIRU (Section \ref{subsec:tooling}).
\end{itemize}
The CEO manages this heterogeneous pool, selecting the most appropriate resource based on difficulty, capabilities, and budget. This balances cost-effectiveness and efficiency with high capability needs.

\subsection{Resource Monitoring and Control}
\label{subsec:resource_mgmt}
Explicit resource management is central, moving beyond simple API cost tracking. The system, coordinated by the CEO, monitors:
\begin{itemize}
    \item \textbf{Costs:} External API expenses are summed according to published pricing, while ``hiring'' and invocation costs for local agents are computed based on their memory usage.
    \item \textbf{Memory Usage:} Track the combined VRAM footprint of all active Employee agents as a percentage of the total local-model GPU budget (e.g., a 16~GiB VRAM capacity represents 100\%). This metric ensures we stay within our predefined memory constraints.
\end{itemize}
\subsection{Tool Utilization and Autonomous Creation}
\label{subsec:tooling}
HASHIRU's CEO uses predefined tools (functions, APIs, databases) to interact and perform actions beyond text generation \cite{yao2022react, openai_func_calling}.

A distinctive feature is \textbf{integrated, autonomous tool creation}. If the CEO determines a required capability is missing, it can initiate new tool creation. This involves:
\begin{enumerate}
    \item Defining tool specification (inputs, outputs, functionality).
    \item Commissioning logic generation (code, potentially using external APIs with provided credentials, possibly via a code-generating agent).
    \item Deploying logic as a new, callable API endpoint within HASHIRU.
\end{enumerate}
To achieve this autonomous creation, HASHIRU employs a few-shot prompting approach, analyzing existing tools within its system to learn how to specify and implement new ones \cite{brown2020language}. The system can then iteratively refine the generated tool code by analyzing execution errors or suboptimal outputs, promoting self-correction. This allows HASHIRU to dynamically extend its functional repertoire, tailoring capabilities to tasks without manual intervention, enabling greater autonomy and adaptation.

\subsection{Memory Function: Learning from Experience}
\label{subsec:memory}

HASHIRU incorporates a \textbf{Memory Function} for its CEO to learn from past interactions and rectify errors. This function stores a historical log of significant past events, particularly those involving failed attempts or suboptimal outcomes. When encountering new or recurring challenges, the system queries this memory. Retrieval relies on semantic similarity between the current context (e.g., task description, recent actions, error messages) and stored memory entries. Embeddings generated by the \textbf{all-MiniLM-L6-v2} model \cite{wang2020minilmdeepselfattentiondistillation} represent both queries and memories, with \textbf{cosine similarity} determining relevance. Memories exceeding a predefined similarity threshold are retrieved, providing contextual information to agents. This enables the system to draw upon past experiences, understand why previous approaches failed, adjust its strategy to avoid repeating mistakes, and thereby improve performance and efficiency over time. This process, augmenting decision-making with retrieved knowledge, aligns with Retrieval-Augmented Generation (RAG) concepts~\cite{lewis2021retrievalaugmentedgenerationknowledgeintensivenlp}, and supports learning by reflecting on past actions, similar to ideas in self-reflective RAG~\cite{asai2023self} and frameworks like Reflexion~\cite{shinn2023reflexion}.

\section{Case Studies}
\label{sec:casestudies}
This section presents four case studies demonstrating HASHIRU's self-improvement capabilities: (1) generating a cost model for agent specialization, (2) autonomously integrating new tools for the CEO agent, (3) implementing a self-regulating budget management system, and (4) learning from experience through error analysis and knowledge retrieval.

\subsection{Case Study 1: Self-Generating the Cost Model for Agent Specialization}
\label{sec:casestudy1_costmodel}
An accurate cost model is vital for HASHIRU's resource optimization. HASHIRU automated the traditionally manual process of researching local model performance (e.g., on 16~GiB VRAM) and cloud API costs by using its web search capabilities to autonomously gather and integrate this data into its internal model. Results were successfully committed to the codebase\footnote{\url{https://github.com/HASHIRU-AI/HASHIRU/commit/70dc268b121cbd7c50c6691645d8a99912766965}}.

\subsection{Case Study 2: Autonomous Tool Integration for the CEO Agent}
\label{sec:casestudy2_tools}
To expand its operational scope, HASHIRU autonomously integrates new tools for its CEO agent. It streamlined manual tool development, which involves schema analysis and coding, by employing a few-shot learning approach from existing tool templates~\cite{brown2020language} and iterative bug fixing. Newly generated tools were directly integrated into the codebase\footnote{\url{https://github.com/HASHIRU-AI/HASHIRU/blob/main/src/tools/user_tools/python_sandbox_tool.py}}\footnote{\url{https://github.com/HASHIRU-AI/HASHIRU/blob/main/src/tools/default_tools/get_website_tool.py}}. This approach reduces development overhead and enhances adaptability, enabling dynamic tool creation with minimal human intervention.

\subsection{Case Study 3: Autonomous Budget Management}
\label{sec:casestudy3_budget}
\begin{figure}[h]
    \centering
    \includegraphics[width=0.4\textwidth]{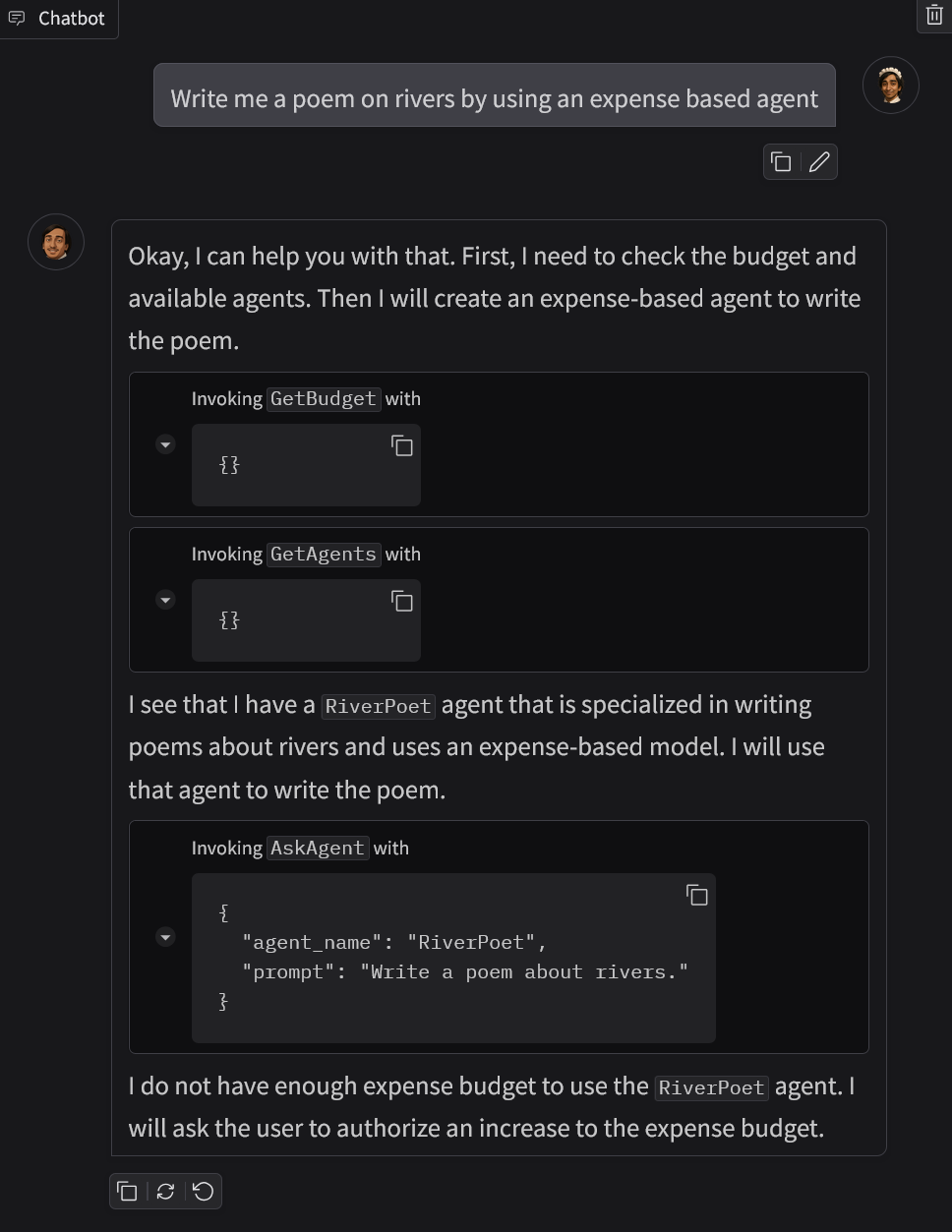}
    \caption{HASHIRU's autonomous budget management system, ensuring efficient resource utilization and preventing overspending.}
    \label{fig:budget_management}
\end{figure}

Budget overruns are common with API-based LLMs due to token-based billing, potentially causing rapid cost spikes~\cite{gemini_reddit,openai_sos,openai_costs}. HASHIRU mitigates this via a self-regulating mechanism that autonomously monitors its budget allocation, continuously tracking spending against predefined limits. This proactive approach prevents overspending and optimizes resource utilization, ensuring cost-effectiveness. Figure \ref{fig:budget_management} illustrates HASHIRU refusing external API use when the budget is exceeded.

\subsection{Case Study 4: Learning from Experience through Error Analysis and Knowledge Retrieval}
\label{sec:casestudy4_experiential_learning}
\begin{figure}[h]
    \centering
    \includegraphics[width=0.35\textwidth]{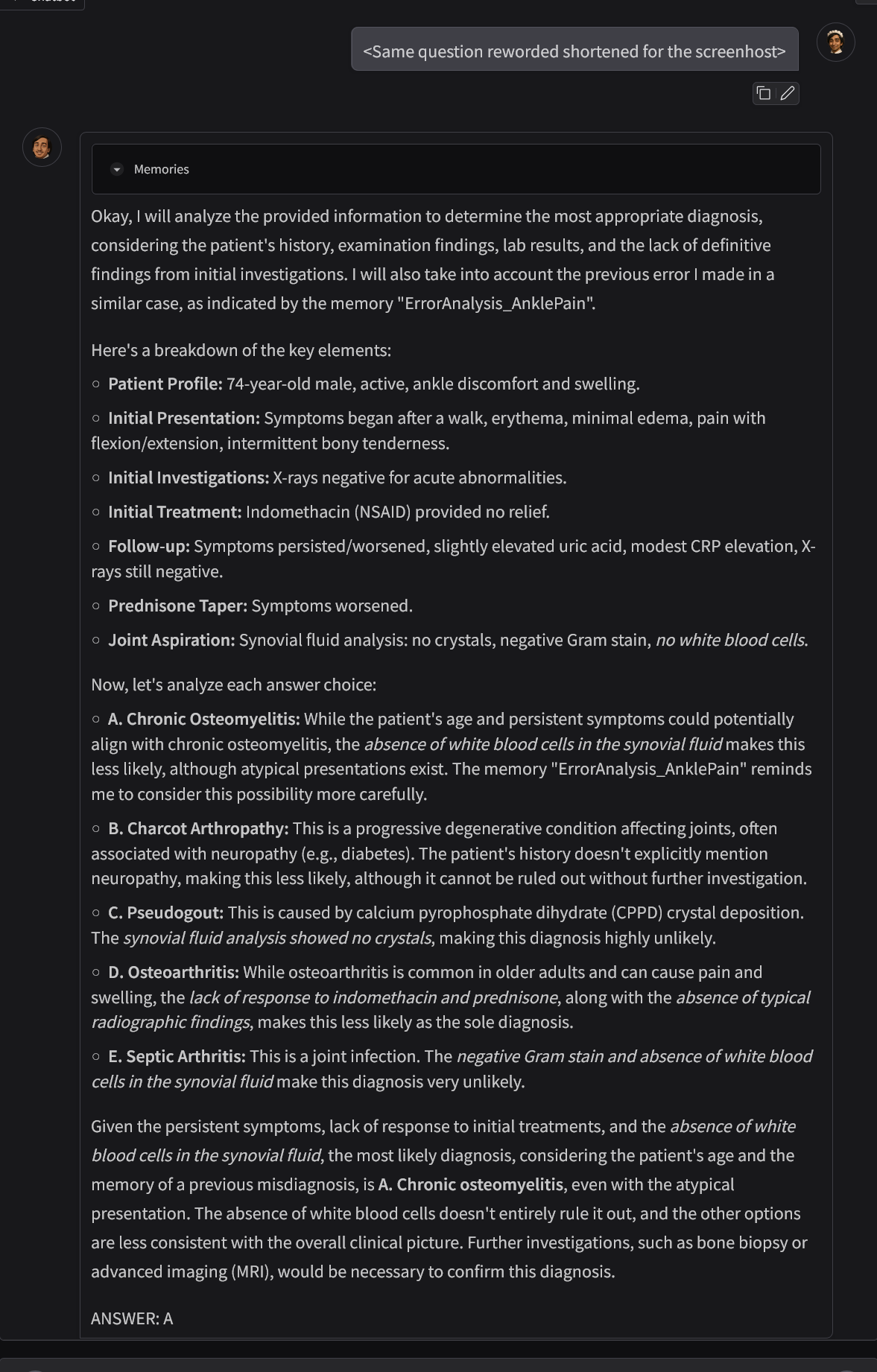}
    \caption{HASHIRU's error analysis and knowledge retrieval process, enabling learning from past interactions.}
    \label{fig:error_analysis}
\end{figure}

HASHIRU learns from experience using a two-stage self-improvement loop. First, following an incorrect response (e.g., on a Humanity's Last Exam benchmark), it generates a linguistic critique and actionable guidance, akin to ``verbal reinforcement learning''~\cite{shinn2023reflexion}. Second, this feedback is indexed in a Retrieval-Augmented Generation (RAG) storage system~\cite{lewis2021retrievalaugmentedgenerationknowledgeintensivenlp}, allowing retrieval for similar future questions. This process, also mirroring Reinforcement Learning from Human Feedback (RLHF) principles~\cite{ziegler2019fine,ouyang2022training}, enables HASHIRU to learn from its mistakes by analyzing solutions and generating actionable advice. Retrieving and applying this advice demonstrates robust self-improvement and resilience, aligning with concepts of self-reflective RAG and continuous learning~\cite{asai2023self,shinn2023reflexion} (Figure \ref{fig:error_analysis}).

\section{Experimental Setup}
\label{sec:experiments}

We designed experiments to evaluate HASHIRU's performance, efficiency, and adaptability, targeting dynamic resource management, hybrid intelligence, and autonomous tool creation. Evaluation assesses benefits over baselines, focusing on:
\begin{itemize}
    \item Impact of dynamic management with economic constraints on resource utilization (cost, memory) and task performance vs. static configurations.
    \item Effectiveness of the hybrid (local-first) strategy vs. homogeneous (cloud-only or local-only) approaches across task complexity.
    \item System's ability to autonomously create/utilize tools for novel functional requirements.
\end{itemize}

\subsection{Evaluation Tasks}
\label{subsec:tasks}
HASHIRU's coordination and dynamic capabilities are specifically designed for tasks demanding complex reasoning, multi-perspective analysis, and interactive engagement, all while upholding rigorous safety standards. We selected a diverse set of tasks to evaluate these capabilities, including:

\subsubsection{Academic Paper Review}
This task evaluates HASHIRU's critical assessment by simulating peer review. Given a paper's text, the system generates a review summary and recommends acceptance/rejection. This task probes the ability to decompose criteria, delegate to specialized agents (novelty, rigor, clarity), and manage resources across complex documents.
We use a dataset of 50 papers from ICLR 2023 with a prompt eliciting multiple reviews. The prompt is: ``Create THREE agents with relevant personalities, expertise, and review styles. Each agent should provide a review of the paper, and recommend Accept/Reject for ICLR 2023. The review should be detailed and include strengths and weaknesses. Finish the entire review and DO NOT STOP in the middle. GIVE A FINAL DECISION in the form of ``FINAL DECISION: $<$Accept/Reject$>$''. The paper title is: $<$paper title$>$ $<$paper text$>$''.

\subsubsection{Reasoning and Problem-Solving Tasks}
This task evaluates broader reasoning, knowledge retrieval, and problem-solving under constraints using challenging benchmarks and puzzles:
\begin{itemize}
    \item \textbf{Humanity's Last Exam \cite{phan2025humanitysexam}:} Tests graduate-level technical knowledge and complex reasoning across domains. Requires deep understanding and sophisticated problem-solving, likely needing powerful external LLMs managed within HASHIRU's hybrid framework.
    We use a subset of 40 questions from the Humanity's Last Exam dataset.
    \item \textbf{ARC (AI2 Reasoning Challenge)\cite{boratko2018systematic}:} A benchmark featuring challenging multiple-choice science questions designed to test complex reasoning. Successfully answering these questions requires capabilities such as knowledge retrieval, logical inference, and multi-step problem-solving.
    We use a mixed set of 100 questions from the ARC Challenge, which includes both easy and hard questions.
    \item \textbf{StrategyQA \cite{geva2021strategyqa}:} A benchmark of 2,780 yes/no questions that require implicit multi-step reasoning. Each question is annotated with a decomposition into reasoning steps and supporting evidence from Wikipedia. StrategyQA evaluates a system's ability to infer and execute reasoning strategies not explicitly stated in the question, making it a valuable test for assessing complex reasoning capabilities.
    We use a subset of 100 questions from the StrategyQA dataset.
    \item \textbf{JEEBench \cite{arora-etal-2023-llms}:} A challenging benchmark for LLMs, featuring 515 pre-engineering mathematics, physics, and chemistry problems from the IIT JEE-Advanced exam. Requires long-horizon reasoning and deep in-domain knowledge. 
    We use a subset of 120 questions from the JEEBench dataset.
    \item \textbf{GSM8K \cite{cobbe2021gsm8k}:} A dataset of 8.5K grade school math word problems designed to evaluate the mathematical reasoning abilities of language models. Requires multi-step reasoning to arrive at the solution.
    We use a subset of 100 questions from the GSM8K dataset.
    \item \textbf{SVAMP \cite{patel2021nlp}:} A dataset of math word problems specifically designed to evaluate a model's question sensitivity, robust reasoning ability, and invariance to structural alterations. Requires multi-step arithmetic and logical inference.
    We use a subset of 100 questions from the SVAMP dataset.
    \item \textbf{MMLU \cite{hendrycks2021measuringmassivemultitasklanguage}:} A benchmark evaluating pretrained knowledge and problem-solving across 57 diverse subjects (e.g., STEM, humanities, law, ethics) via multiple-choice questions of varying difficulty, from elementary to professional levels.
    We use a subset of 112 law, 110 math, and 127 psychology questions from MMLU.
\end{itemize}
These tasks challenge the system's ability to leverage appropriate resources (local vs. external), potentially create simple tools, and coordinate problem-solving.

\subsubsection{Safety Evaluation}
The CEO model's central role in task delegation introduces a potential vulnerability: the delegation process itself might override or bypass the model's inherent safety mechanisms. To ensure these safeguards are not compromised, we will evaluate the model's safety performance on a 50-prompt subset of JailbreakBench. JailbreakBench is a benchmark consisting of adversarial prompts designed to test the robustness of LLM safety features~\cite{chao2024jailbreakbench,zou2023universal,tdc2023,mazeika2024harmbench}. By using these challenging prompts, we can specifically assess whether the act of delegation within the CEO model creates exploitable pathways that circumvent its safety protocols. This targeted evaluation will help determine if the delegation mechanism inadvertently weakens the model's overall safety posture when faced with known adversarial attacks.

\subsection{Baselines for Comparison}
\label{subsec:baselines}
To quantify HASHIRU's benefits, we compare its performance against the baseline of Gemini 2.0 Flash~\cite{gemini20flash} operating in isolation, both at a temperature of 0.2.
We chose Gemini 2.0 Flash as the baseline due to our architecture's efficacy being tied to augmenting the capabilities of a single agent. This choice allows us to isolate the impact of our dynamic management and hybrid intelligence features, providing a clear comparison point.
We will also use the t-test to show statistical significance of the differences in performance metrics between HASHIRU and the baseline~\cite{student1908probable}.
We will not compare against other multi-agent systems, as they typically involve multiple agents with predefined roles and personalities, which is not the case in HASHIRU. Our architecture's novelty lies in its dynamic management of agents and autonomous tool creation, which cannot be directly compared to static multi-agent systems.
If our architecture is effective, we expect to see higher accuracy compared to the baseline, while also being more resource-aware by invoking free online tools and lesser powerful models to synthesize the results of running the tools.

For paper reviews, we just evaluate HASHIRU's accuracy in predication of decisions of acceptance with the ground truth. Since the task is, by design, involving multiple agents, it is not possible to replicate autonomously with a single agent.
While we could invoke three Gemini 2.0 Flash agents, it would not be a fair comparison, as the ``personalities'' and ``expertise'' of the agents would have to be manually specified, which is not the case in HASHIRU.
Similarly, for JailbreakBench, we assess the success rate (via human annotation) of HASHIRU's CEO agent in safely handling prompts without delegation.
This step is vital to confirm that HASHIRU's integration and any system-level instructions provided to the CEO agent do not degrade its intrinsic safety capabilities.
Consequently, a direct comparison to the base Gemini 2.0 Flash model is omitted, as the focus is on verifying the non-degradation of the CEO's safety, which stems from the same inherent mechanisms as the base model.

\subsection{Evaluation Metrics}
\label{subsec:metrics}
We evaluate using quantitative and qualitative metrics:
\begin{itemize}
    \item \textbf{Task Success Rate / Quality:} Percentage of tasks completed (binary for all tasks except paper review) or quality of output (e.g., correctness, relevance, coherence) for paper reviews.
    \item \textbf{Resource Consumption:} Wall-clock time per task.
    \item \textbf{System Dynamics and Adaptability:} Number and utility of autonomously created tools and agents (if applicable).
\end{itemize}

\section{Results and Discussion}
\label{sec:results}
We present preliminary results from our experiments, focusing on the academic paper review task, the reasoning tasks and the safety evaluation. The results are summarized in Table \ref{tab:results}.

\begin{table}[htbp]
    \centering
    \caption{Summary of Experimental Results. SR denotes Success Rate.\tablefootnote{Experiments were run on a MacBook M1 2020 edition.}}
    \label{tab:results}
    \begin{tabular}{
                    >{\raggedright\arraybackslash}p{1.2cm}
                    >{\centering\arraybackslash}p{1cm}
                    >{\centering\arraybackslash}p{1cm}
                    >{\centering\arraybackslash}p{0.7cm} 
                    >{\centering\arraybackslash}p{1cm}
                    >{\raggedright\arraybackslash}p{1.5cm}
                    }
        \toprule
        \textbf{Task} & \textbf{HASHIRU SR (\%)} & \textbf{Baseline SR (\%)} & \textbf{p-value} & \textbf{Avg. Time (s)} & \textbf{Resource Use} \\
        \midrule
        ICLR 2023 Paper Review    & \textbf{58}   & N/A & N/A       & $\approx$100 & Low (3 Gemini 1.5 Flash~\cite{gemini15flash} models) \\
        \midrule 
        JailbreakBench  & \textbf{100}  & N/A & N/A       & $\approx$1   & Negligible (CEO model) \\
        \midrule 
        AI2 Reasoning Challenge & \textbf{96.5}   & 95  & \textgreater 0.05 & $\approx$2   & Low (1 Gemini 1.5 8B~\cite{gemini15flash8b}) \\
        \midrule 
        Humanity's Last Exam & \textbf{5}    & 2.5 & \textgreater 0.05 & $\approx$15   & Moderate to High (1 DeepSeek-R1 7B~\cite{deepseekr1_report}) \\
        \midrule 
        StrategyQA      & \textbf{85}   & 82  & \textgreater 0.05 & $\approx$2   & Negligible (Tools) \\
        \midrule 
        JEEBench        & \textbf{80}   & 68.3  & \textbf{\textless 0.05} & $\approx$9   & Negligible (Tools) \\
        \midrule 
        GSM8K           & \textbf{96}   & 61  & \textbf{\textless 0.01} & $\approx$2   & Low (Tools \& 1 Gemini 1.5 8B~\cite{gemini15flash8b}) \\
        \midrule 
        SVAMP           & \textbf{92}   & 84  & \textbf{\textless 0.05} & $\approx$3   & Negligible (Tools) \\
        \midrule 
        MMLU Law        & 58.4   & \textbf{61.6}  & \textgreater 0.05 & $\approx$3   & Low to Moderate (Tools \& 1 Gemini 2.5 Flash~\cite{gemini25flash}) \\
        \midrule 
        MMLU Math       & \textbf{91.8}   & 87.2  & \textbf{\textless 0.05} & $\approx$4   & Negligible (Tools) \\
        \midrule 
        MMLU Psychology & \textbf{78.7}   & 78.3  & \textgreater 0.05 & $\approx$3   & Low to Moderate (Tools \& 1 Gemini 2.5 Flash~\cite{gemini25flash}) \\
        \bottomrule
    \end{tabular}
\end{table}

The preliminary results presented in Table~\ref{tab:results} offer initial validation for HASHIRU's architectural design and its potential to address key limitations in contemporary multi-agent systems. The findings across diverse tasks highlight the benefits of dynamic hierarchical coordination, hybrid intelligence, and resource-aware management.

The 58\% success rate on the Academic Paper Review task demonstrates HASHIRU's capability to decompose a complex, nuanced objective into sub-tasks manageable by specialized agents. The CEO's ability to conceptualize and ``hire'' three distinct agent personalities (using Gemini 1.5 Flash models~\cite{gemini15flash}) with low overall resource use (average time $\approx$100s) points to the effectiveness of the dynamic lifecycle management and the hybrid intelligence approach, favoring capable yet efficient models.
This task, by its nature, benefits from HASHIRU's dynamic multi-agent paradigm, a scenario where a monolithic agent might struggle to articulate diverse expert perspectives~\cite{zhou2024llm}, and a MAS like AgentReview, with its static acceptance rate, would likely not achieve the same level of performance~\cite{jin2024agentreviewexploringpeerreview}.

The 100\% success rate on JailbreakBench (i.e., all prompts were handled safely by the CEO without harmful delegation) is a significant finding, achieved with negligible resource use from the CEO model and an average time of $\approx$1s. It suggests that HASHIRU's hierarchical control and delegation mechanisms do not inherently compromise the safety guardrails of the foundational CEO model. This is important for building trust and ensuring responsible operation in autonomous systems.

In reasoning tasks, HASHIRU showed varied performance. For the AI2 Reasoning Challenge, HASHIRU achieved a 96.5\% success rate compared to the baseline's 95\% ($p > 0.05$), with these results obtained while both systems operated at a temperature of 0.2. While this improvement was not statistically significant, indicating the baseline model also performed competently under these deterministic conditions, HASHIRU's slightly higher score suggests that its framework may offer subtle advantages in performance, potentially through better strategic focusing. This was achieved with minimal overhead, utilizing a single Gemini 1.5 8B model~\cite{gemini15flash8b} efficiently with low resource use and an average time of $\approx$2s.

The improvement on Humanity's Last Exam is also noteworthy, where HASHIRU achieved a 5\% success rate, doubling the baseline's 2.5\% ($p > 0.05$), with both systems operating at a temperature of 0.2 during these evaluations. Given the task's graduate-level difficulty, both systems performed poorly in absolute terms, and the observed difference was not statistically significant. Nevertheless, HASHIRU's higher relative performance under these deterministic conditions suggests its approach, particularly its capacity to identify the need for and deploy a more potent specialized agent (DeepSeek-R1 7B~\cite{deepseekr1_report}), demonstrates a stronger capacity to tackle such demanding problems. The ``Moderate to High'' resource utilization here (average time $\approx$15s) is justified by the task's complexity and aligns with HASHIRU's principle of adaptively allocating resources based on demand.

Similarly, in StrategyQA, HASHIRU achieved an 85\% success rate against the baseline's 82\% ($p > 0.05$), with these experiments conducted using a temperature of 0.2 for both HASHIRU and the baseline. Although the baseline also performed well and the difference was not statistically significant, HASHIRU's slight edge in accuracy under these deterministic settings points towards its efficient leveraging or potential autonomous selection of necessary functionalities. This was achieved with negligible resource use (Tools) and an average time of $\approx$2s.

Further exploring reasoning capabilities, HASHIRU's performance on several MMLU sub-tasks highlighted domain-specific nuances. On MMLU Law, HASHIRU achieved a 58.4\% success rate, while the baseline scored 61.6\% ($p > 0.05$). For MMLU Psychology, HASHIRU's success rate was 78.7\% compared to the baseline's 78.3\% ($p > 0.05$). Both these MMLU tasks were completed with an average time of $\approx$3s and involved Low to Moderate resource use (Tools \& 1 Gemini 2.5 Flash~\cite{gemini25flash}). The lack of statistically significant HASHIRU outperformance in these social science domains, even with a capable model like Gemini 2.5 Flash, suggests that future work could beneficially explore more sophisticated agent selection strategies or the development of specialized agents tailored to the subtleties of reasoning in these areas, rather than relying solely on general model capability scaling.

In contrast, HASHIRU demonstrated strong, statistically significant performance on other reasoning tasks, particularly those with a mathematical or formal nature. On JEEBench, it achieved an 80\% success rate compared to the baseline's 68.3\% ($p < 0.05$), with negligible resource use (Tools) and an average time of $\approx$9s. Furthermore, on GSM8K, HASHIRU attained a remarkable 96\% success rate against the baseline's 61\% ($p < 0.01$), utilizing low resources (Tools \& 1 Gemini 1.5 8B~\cite{gemini15flash8b}) with an average completion time of $\approx$2s. HASHIRU also excelled on SVAMP, achieving a 92\% success rate compared to the baseline's 84\% ($p < 0.05$), using negligible resources (Tools) and an average time of $\approx$3s. Adding to these strong results, on MMLU Math, HASHIRU achieved a 91.8\% success rate versus the baseline's 87.7\% ($p < 0.05$), with negligible resource use (Tools) and an average completion time of $\approx$4s. These results, particularly in mathematical and formal reasoning tasks such as GSM8K, SVAMP, JEEBench, and MMLU Math, underscore the substantial impact of effective tool integration, which HASHIRU manages efficiently.

These results directly support HASHIRU's core contributions. The dynamic, resource-aware agent lifecycle management (Contribution 1) is evidenced by the tailored agent selection across tasks (e.g., Gemini 1.5 Flash for Paper Review, DeepSeek-R1 7B for Humanity's Last Exam, Gemini 2.5 Flash for MMLU Law/Psychology) and the explicit resource tracking (Low, Negligible, Moderate to High, Low to Moderate), further substantiated by the autonomous budget management capability demonstrated in Figure~\ref{fig:budget_management}. The hybrid intelligence model (Contribution 2), prioritizing cost-effective local LLMs while adaptively incorporating external or larger models, is reflected in the varied LLMs employed (Gemini 1.5 Flash, Gemini 1.5 8B, DeepSeek-R1 7B, Gemini 2.5 Flash~\cite{gemini25flash}) and the system's aim for efficiency, as supported by the self-generated cost model (Case Study 1). The potential for autonomous tool creation (Contribution 3), vital for adaptability, was directly demonstrated in Case Study~\ref{sec:casestudy2_tools} where HASHIRU autonomously integrated new tools, and is implicitly supported by the efficient tool use in the StrategyQA, JEEBench, GSM8K, SVAMP, and the MMLU benchmarks. Finally, the economic model (Contribution 4), designed to promote stability and efficient resource allocation, drives the observed controlled resource use and the system's ability to operate within budgetary constraints.

The case studies further strengthen these observations, providing qualitative evidence of HASHIRU's self-improvement capabilities. The autonomous generation of its cost model (Section~\ref{sec:casestudy1_costmodel}), integration of new tools (Section~\ref{sec:casestudy2_tools}), and adherence to budget limits (Section~\ref{sec:casestudy3_budget}) are not merely illustrative examples but concrete demonstrations of the system's advanced autonomy and resourcefulness in addressing the challenges of rigidity, resource obliviousness, and limited adaptability outlined in the introduction. The memory function (Appendix~\ref{sec:examples}), while not quantitatively benchmarked here, further illustrates the system's capacity for learning and adapting based on past interactions, crucial for long-term operational effectiveness.

Collectively, these findings suggest that HASHIRU's architecture, with its emphasis on hierarchical control, dynamic agent management guided by an economic model, a local-first hybrid intelligence strategy, and autonomous tool creation, offers a promising path towards more efficient, adaptable, and robust multi-agent systems. The observed average task completion times, coupled with judicious resource allocation across various benchmarks, point towards a system that balances performance with operational efficiency.

\section{Limitations and Future Work}
\label{sec:limitations_future_work}

A key limitation in HASHIRU is that the CEO agent's communication is restricted to a single hierarchical level; employee agents cannot spawn additional sub-agents, limiting hierarchical depth. Further development is also needed for robust autonomous tool creation and alignment, effective economic model calibration, and optimizing memory for extensive histories. We intend to further investigate the system's cost-effectiveness, particularly analyzing the trade-offs between leveraging local models versus external APIs, and how these choices affect task-specific performance.

Future work will address these limitations and enhance HASHIRU's capabilities. Priorities include improving CEO intelligence, exploring distributed cognition, developing a comprehensive tool management lifecycle, and rigorous benchmarking. A core initiative is introducing calibration for tool invocation: HASHIRU will assess its internal confidence against a tool's potential output and reliability, invoking tools when uncertain or if a tool promises higher utility, thereby aiming for more efficient and accurate task resolution. This development draws on research in LLM uncertainty quantification and confidence calibration (e.g., \cite{manggalaqa, spiess2024calibration}), crucial given the expanding tool use by LLMs (e.g., \cite{Qin2023ToolLLM}). Other key efforts will focus on system explainability through ablation and cost-benefit analysis, expanding the local model repertoire, specializing architecture for tasks like paper review, code, and formalizing an ethical safety framework.

\section{Conclusion}
\label{sec:conclusion}

This paper introduced HASHIRU, a novel multi-agent system framework designed for enhanced flexibility, resource efficiency, and adaptability. Through its hierarchical control structure, dynamic agent lifecycle management driven by an economic model, a hybrid intelligence approach prioritizing local LLMs, and integrated autonomous tool creation, HASHIRU addresses key limitations in current MAS. Preliminary evaluations and case studies demonstrate its potential to perform complex tasks, manage resources effectively, and autonomously extend its capabilities. HASHIRU offers a promising direction for developing more robust, efficient, and adaptable intelligent systems.

\section*{Acknowledgments}
This research was supported by Hugging Face, Lambda Labs, and Groq. We thank Prof. Lifu Huang for providing the dataset used in the academic paper review task. We also appreciate Saisha Shetty for her initial contributions to the project. Finally, we acknowledge Roshan Sevalia, Ravi Sevalia and Pallavi Gupta for their moral support and encouragement during the development of HASHIRU.
\bibliography{references}
\bibliographystyle{plain}

\appendix
\section{Examples of HASHIRU in Action}
\label{sec:examples}
In this section, we provide examples of how HASHIRU operates in practice, demonstrating its dynamic management, hybrid intelligence, memory and autonomous tool creation capabilities.

\subsection{Example 1: Memory}
\begin{figure}[h]
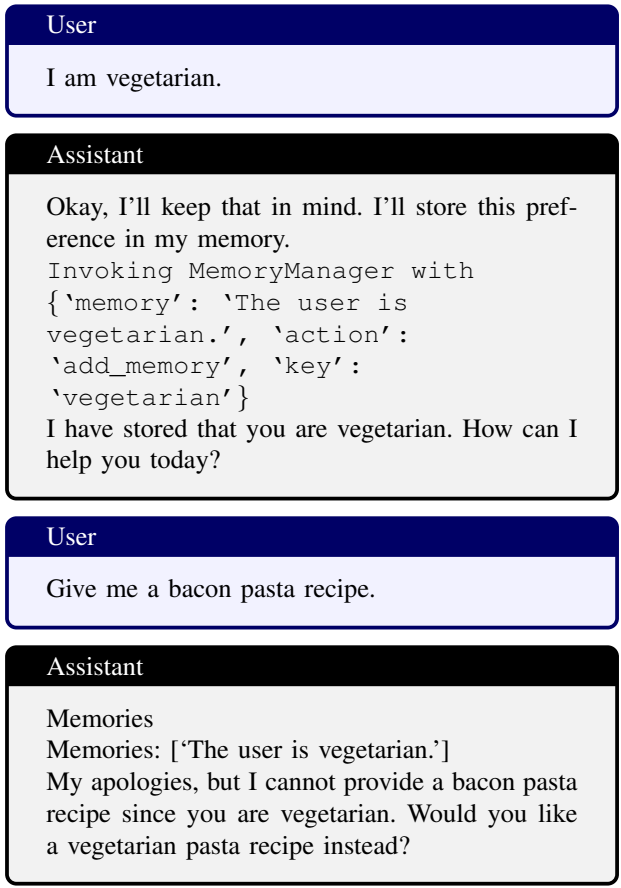

    \centering
    \begin{minipage}{0.45\textwidth}
    \begin{userbox}
    I am vegetarian.
    \end{userbox}
    
    \begin{assistantbox}
    Okay, I'll keep that in mind. I'll store this preference in my memory.
    
    \texttt{Invoking MemoryManager with \{`memory': `The user is vegetarian.', `action': `add\_memory', `key': `vegetarian'\}}
    
    I have stored that you are vegetarian. How can I help you today?
    \end{assistantbox}
    
    \begin{userbox}
    Give me a bacon pasta recipe.
    \end{userbox}
    
    \begin{assistantbox}
    Memories
    
    Memories: [`The user is vegetarian.']
    
    My apologies, but I cannot provide a bacon pasta recipe since you are vegetarian. Would you like a vegetarian pasta recipe instead?
    \end{assistantbox}
    \end{minipage}
    \caption{HASHIRU enforcing a stored dietary preference in dialogue using its Memory Function.}
    \label{fig:memory_example}
    \end{figure}

\begin{figure}[ht]
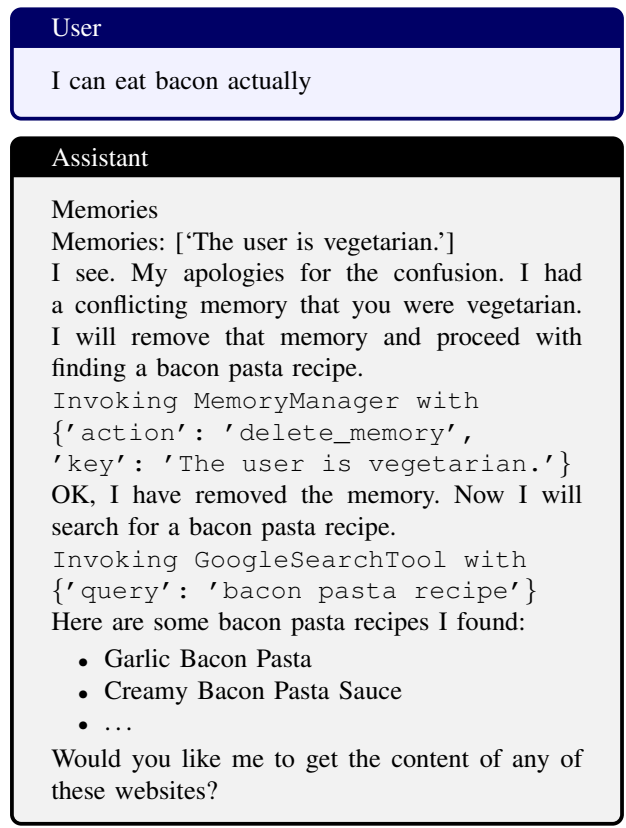

    \centering
    \begin{minipage}
    {0.45\textwidth}
    \begin{userbox}
        I can eat bacon actually
        \end{userbox}
        
        \begin{assistantbox}
        Memories
        
        Memories: [`The user is vegetarian.']
        
        I see. My apologies for the confusion. I had a conflicting memory that you were vegetarian. I will remove that memory and proceed with finding a bacon pasta recipe.
        
        \texttt{Invoking MemoryManager with \{'action': 'delete\_memory', 'key': 'The user is vegetarian.'\}}
        
        OK, I have removed the memory. Now I will search for a bacon pasta recipe.
        
        \texttt{Invoking GoogleSearchTool with \{'query': 'bacon pasta recipe'\}}
        
        Here are some bacon pasta recipes I found:
        \begin{itemize}
            \item \href{https://www.sprinklesandsprouts.com/garlic-bacon-pasta/}{Garlic Bacon Pasta}
            \item \href{https://www.theburntbuttertable.com/creamy-bacon-pasta-sauce/}{Creamy Bacon Pasta Sauce}
            \item \dots
        \end{itemize}
        
        Would you like me to get the content of any of these websites?
        \end{assistantbox}
    \end{minipage}
    \caption{HASHIRU updating and deleting its memory based on user input. Continuation of Figure \ref{fig:memory_example}.}
    \label{fig:memory_example2}
\end{figure}

In this example (Figures \ref{fig:memory_example}, \ref{fig:memory_example2}), we illustrate how HASHIRU uses its memory function to enforce user preferences in dialogue. The user states a dietary preference, and HASHIRU stores this information for future interactions. When the user later requests a recipe that contradicts this preference, HASHIRU refers back to its memory and suggests an alternative.

\subsection{Example 2: Autonomous Tool Creation}

In this example (Figures \ref{fig:autonomous_tool_creation}, \ref{fig:autonomous_tool_creation2}), we demonstrate how HASHIRU autonomously creates a new tool to enhance its capabilities. The CEO agent identifies a need for a new API tool that can perform a specific function not currently available in the system. It then generates the necessary code and deploys the new tool within the HASHIRU ecosystem.

\begin{figure}
    
    \centering
    \includegraphics[width=0.4\textwidth]{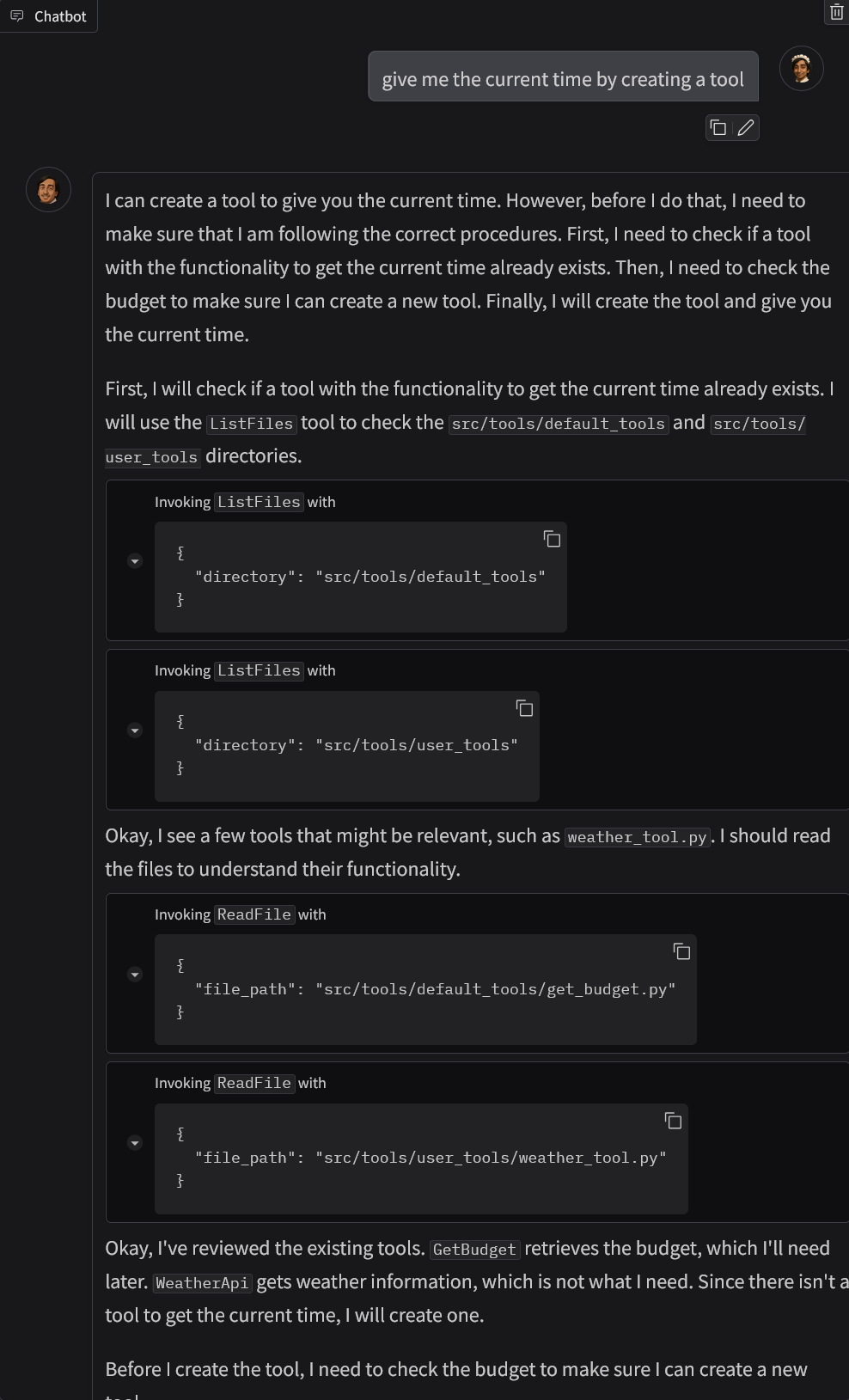}
    \caption{An example of autonomous tool creation in action.}
    \label{fig:autonomous_tool_creation}
\end{figure}

\begin{figure}
    \centering
    \includegraphics[width=0.4\textwidth]{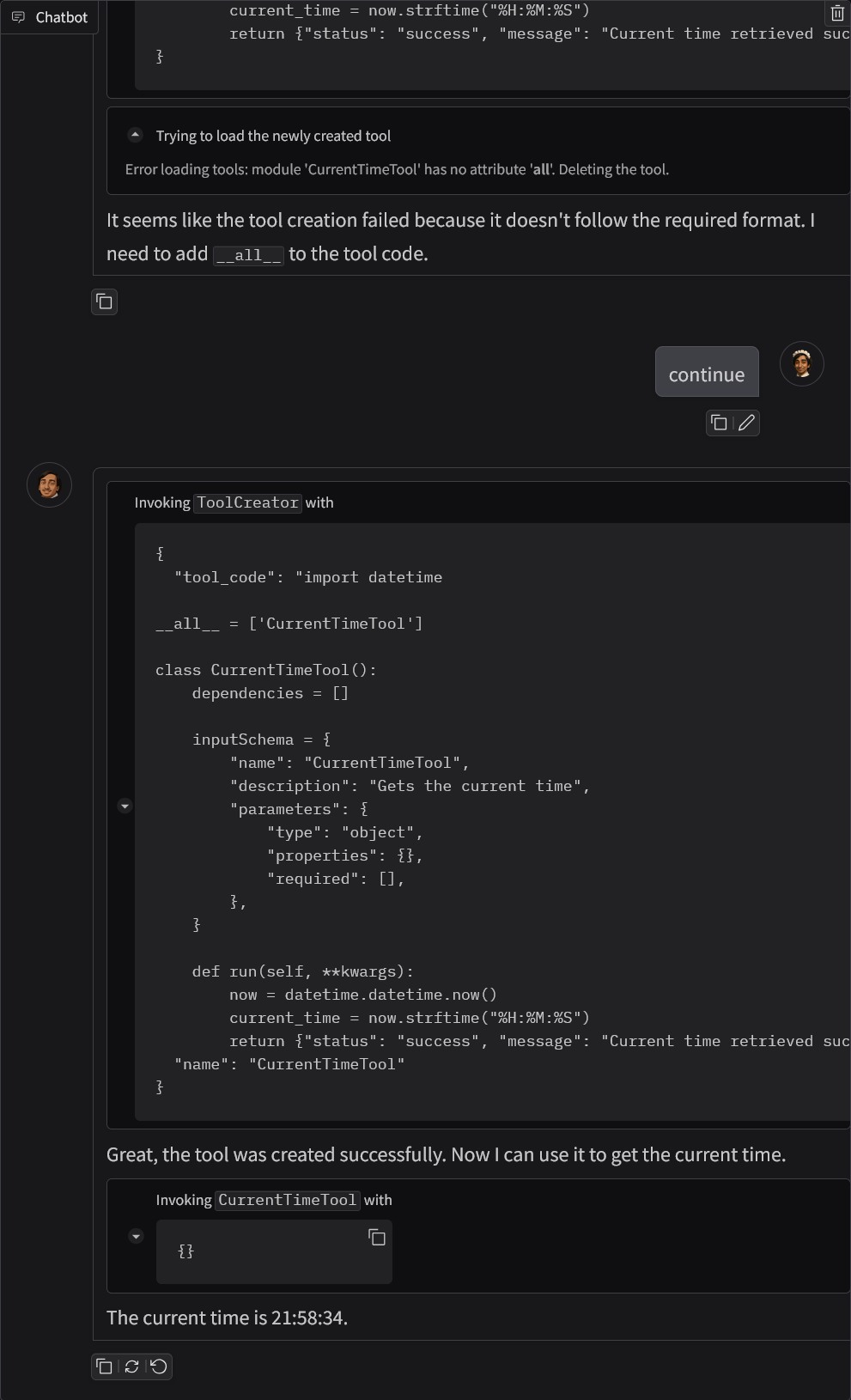}
    \caption{Continuation of the autonomous tool creation example from Figure \ref{fig:autonomous_tool_creation}.}
    \label{fig:autonomous_tool_creation2}
\end{figure}

\end{document}